\documentclass[12pt,a4paper]{article}
\usepackage{epsfig}
\newcommand{\lumi}{\mbox{$\mathrm{pb}^{-1}\,$}}
\newcommand{\mlumi}{\,{\rm pb}^{-1}\,}
\newcommand{\GeV}{GeV}

\newcommand{\GeVc}{$\mathrm{GeV}/c$}
\newcommand{\Wvis}{$W_{\mathrm{vis}}$}
\newcommand{\GeVcsq}{$\mathrm{GeV/}c^2$\hspace{0.2ex}}
\newcommand{\pclus}{p_{\mbox{\unboldmath$\mathrm{jet}$}}}

\newcommand{\xgmin}{$x_{\gamma}^{\mathrm{min}}\,$}

\newcommand{\beauty}{b}

\newcommand{\Deye}{$D_i$}
\newcommand{\Done}{$D_1$}
\newcommand{\Dtwo}{$D_2$}
\newcommand{\Dthree}{$D_3$}
\newcommand{\mathDthree}{\mathnormal{D}_3}
\newcommand{\tagged}{$\gamma^* \gamma\,$}

\begin{document}
\thispagestyle{empty}
\large{EUROPEAN ORGANISATION FOR NUCLEAR RESEARCH (CERN)}

\vspace*{1.5cm}

\begin{flushright}
{\small CERN-EP/2007-xxx} \\
{\small 21 June 2007} \\
\end{flushright}

\vspace{20mm}

\begin{center}
\LARGE \bf Measurement of the Cross Section \\ for  open b-Quark
Production \\in Two-Photon Interactions at LEP
\end{center}
\vspace*{10mm}
\renewcommand{\thefootnote}{*}
\begin{center}
{ The ALEPH Collaboration
\footnote{See next page for the list of authors}
}
\vspace{5mm} \\
\end{center}
\begin{center}
{\bf Abstract}
\end{center}

\noindent
Inclusive \beauty-quark production in two-photon collisions has been measured at LEP using
an integrated luminosity of
$698~\lumi$  collected by the ALEPH detector with $\sqrt{s}$
between 130 and 209 \GeV . The b quarks were identified using lifetime information.
 The cross section is found to be \[ \mathrm{ \sigma(e^+ e^-
\rightarrow e^+ e^- b \bar{b}\, X)  = (
5.4
\pm
0.8
\,_{stat} \pm
0.8
\,_{syst}} )
\,\mathrm{pb},\]
 which is consistent with  Next-to-Leading Order QCD.
\vspace{10mm} \\
\vspace{\fill}
\begin{center}
{\large \em Submitted to JHEP.}
\noindent
\end{center}
\pagestyle{empty}
\newpage
\small
%
%
\newlength{\saveparskip}
\newlength{\savetopsep}
\setlength{\saveparskip}{\parskip}
\setlength{\savetopsep}{\topsep}
%
%
 \setlength{\parskip}{0.0cm}
 \setlength{\topsep}{1mm}
\pretolerance=10000
\centerline{\large\bf The ALEPH Collaboration}
\footnotesize
\vspace{0.5cm}
{\raggedbottom
\begin{sloppypar}
\samepage\noindent
S.~Schael
\nopagebreak
\begin{center}
\parbox{15.5cm}{\sl\samepage
Physikalisches Institut der RWTH-Aachen, D-52056 Aachen, Germany}
\end{center}\end{sloppypar}
\vspace{2mm}
\begin{sloppypar}
\noindent
R.~Barate,
R.~Bruneli\`ere,
I.~De~Bonis,
D.~Decamp,
C.~Goy,
S.~J\'ez\'equel,
J.-P.~Lees,
F.~Martin,
E.~Merle,
\mbox{M.-N.~Minard},
B.~Pietrzyk,
B.~Trocm\'e
\nopagebreak
\begin{center}
\parbox{15.5cm}{\sl\samepage
Laboratoire de Physique des Particules (LAPP), IN$^{2}$P$^{3}$-CNRS,
F-74019 Annecy-le-Vieux Cedex, France}
\end{center}\end{sloppypar}
\vspace{2mm}
\begin{sloppypar}
\noindent
S.~Bravo,
M.P.~Casado,
M.~Chmeissani,
J.M.~Crespo,
E.~Fernandez,
M.~Fernandez-Bosman,
Ll.~Garrido,$^{15}$
M.~Martinez,
A.~Pacheco,
H.~Ruiz
\nopagebreak
\begin{center}
\parbox{15.5cm}{\sl\samepage
Institut de F\'{i}sica d'Altes Energies, Universitat Aut\`{o}noma
de Barcelona, E-08193 Bellaterra (Barcelona), Spain$^{7}$}
\end{center}\end{sloppypar}
\vspace{2mm}
\begin{sloppypar}
\noindent
A.~Colaleo,
D.~Creanza,
N.~De~Filippis,
M.~de~Palma,
G.~Iaselli,
G.~Maggi,
M.~Maggi,
S.~Nuzzo,
A.~Ranieri,
G.~Raso,$^{24}$
F.~Ruggieri,
G.~Selvaggi,
L.~Silvestris,
P.~Tempesta,
A.~Tricomi,$^{3}$
G.~Zito
\nopagebreak
\begin{center}
\parbox{15.5cm}{\sl\samepage
Dipartimento di Fisica, INFN Sezione di Bari, I-70126 Bari, Italy}
\end{center}\end{sloppypar}
\vspace{2mm}
\begin{sloppypar}
\noindent
X.~Huang,
J.~Lin,
Q. Ouyang,
T.~Wang,
Y.~Xie,
R.~Xu,
S.~Xue,
J.~Zhang,
L.~Zhang,
W.~Zhao
\nopagebreak
\begin{center}
\parbox{15.5cm}{\sl\samepage
Institute of High Energy Physics, Academia Sinica, Beijing, The People's
Republic of China$^{8}$}
\end{center}\end{sloppypar}
\vspace{2mm}
\begin{sloppypar}
\noindent
D.~Abbaneo,
T.~Barklow,$^{26}$
O.~Buchm\"uller,$^{26}$
M.~Cattaneo,
B.~Clerbaux,$^{23}$
H.~Drevermann,
R.W.~Forty,
M.~Frank,
F.~Gianotti,
J.B.~Hansen,
J.~Harvey,
D.E.~Hutchcroft,$^{30}$,
P.~Janot,
B.~Jost,
M.~Kado,$^{2}$
P.~Mato,
A.~Moutoussi,
F.~Ranjard,
L.~Rolandi,
D.~Schlatter,
F.~Teubert,
A.~Valassi,
I.~Videau
\nopagebreak
\begin{center}
\parbox{15.5cm}{\sl\samepage
European Laboratory for Particle Physics (CERN), CH-1211 Geneva 23,
Switzerland}
\end{center}\end{sloppypar}
\vspace{2mm}
\begin{sloppypar}
\noindent
F.~Badaud,
S.~Dessagne,
A.~Falvard,$^{20}$
D.~Fayolle,
P.~Gay,
J.~Jousset,
B.~Michel,
S.~Monteil,
D.~Pallin,
J.M.~Pascolo,
P.~Perret
\nopagebreak
\begin{center}
\parbox{15.5cm}{\sl\samepage
Laboratoire de Physique Corpusculaire, Universit\'e Blaise Pascal,
IN$^{2}$P$^{3}$-CNRS, Clermont-Ferrand, F-63177 Aubi\`{e}re, France}
\end{center}\end{sloppypar}
\vspace{2mm}
\begin{sloppypar}
\noindent
J.D.~Hansen,
J.R.~Hansen,
P.H.~Hansen,
A.C.~Kraan,
B.S.~Nilsson
\nopagebreak
\begin{center}
\parbox{15.5cm}{\sl\samepage
Niels Bohr Institute, 2100 Copenhagen, DK-Denmark$^{9}$}
\end{center}\end{sloppypar}
\vspace{2mm}
\begin{sloppypar}
\noindent
A.~Kyriakis,
C.~Markou,
E.~Simopoulou,
A.~Vayaki,
K.~Zachariadou
\nopagebreak
\begin{center}
\parbox{15.5cm}{\sl\samepage
Nuclear Research Center Demokritos (NRCD), GR-15310 Attiki, Greece}
\end{center}\end{sloppypar}
\vspace{2mm}
\begin{sloppypar}
\noindent
A.~Blondel,$^{12}$
\mbox{J.-C.~Brient},
F.~Machefert,
A.~Roug\'{e},
H.~Videau
\nopagebreak
\begin{center}
\parbox{15.5cm}{\sl\samepage
Laoratoire Leprince-Ringuet, Ecole
Polytechnique, IN$^{2}$P$^{3}$-CNRS, \mbox{F-91128} Palaiseau Cedex, France}
\end{center}\end{sloppypar}
\vspace{2mm}
\begin{sloppypar}
\noindent
V.~Ciulli,
E.~Focardi,
G.~Parrini
\nopagebreak
\begin{center}
\parbox{15.5cm}{\sl\samepage
Dipartimento di Fisica, Universit\`a di Firenze, INFN Sezione di Firenze,
I-50125 Firenze, Italy}
\end{center}\end{sloppypar}
\vspace{2mm}
\begin{sloppypar}
\noindent
A.~Antonelli,
M.~Antonelli,
G.~Bencivenni,
F.~Bossi,
G.~Capon,
F.~Cerutti,
V.~Chiarella,
P.~Laurelli,
G.~Mannocchi,$^{5}$
G.P.~Murtas,
L.~Passalacqua
\nopagebreak
\begin{center}
\parbox{15.5cm}{\sl\samepage
Laboratori Nazionali dell'INFN (LNF-INFN), I-00044 Frascati, Italy}
\end{center}\end{sloppypar}
\vspace{2mm}
\begin{sloppypar}
\noindent
J.~Kennedy,
J.G.~Lynch,
P.~Negus,
V.~O'Shea,
A.S.~Thompson
\nopagebreak
\begin{center}
\parbox{15.5cm}{\sl\samepage
Department of Physics and Astronomy, University of Glasgow, Glasgow G12
8QQ,United Kingdom$^{10}$}
\end{center}\end{sloppypar}
\vspace{2mm}
\begin{sloppypar}
\noindent
S.~Wasserbaech
\nopagebreak
\begin{center}
\parbox{15.5cm}{\sl\samepage
Utah Valley State College, Orem, UT 84058, U.S.A.}
\end{center}\end{sloppypar}
\vspace{2mm}
\begin{sloppypar}
\noindent
R.~Cavanaugh,$^{4}$
S.~Dhamotharan,$^{21}$
C.~Geweniger,
P.~Hanke,
V.~Hepp,
E.E.~Kluge,
A.~Putzer,
H.~Stenzel,
K.~Tittel,
M.~Wunsch$^{19}$
\nopagebreak
\begin{center}
\parbox{15.5cm}{\sl\samepage
Kirchhoff-Institut f\"ur Physik, Universit\"at Heidelberg, D-69120
Heidelberg, Germany$^{16}$}
\end{center}\end{sloppypar}
\vspace{2mm}
\begin{sloppypar}
\noindent
R.~Beuselinck,
W.~Cameron,
G.~Davies,
P.J.~Dornan,
M.~Girone,$^{1}$
N.~Marinelli,
J.~Nowell,
S.A.~Rutherford,
J.K.~Sedgbeer,
J.C.~Thompson,$^{14}$
R.~White
\nopagebreak
\begin{center}
\parbox{15.5cm}{\sl\samepage
Department of Physics, Imperial College, London SW7 2BZ,
United Kingdom$^{10}$}
\end{center}\end{sloppypar}
\vspace{2mm}
\begin{sloppypar}
\noindent
V.M.~Ghete,
P.~Girtler,
E.~Kneringer,
D.~Kuhn,
G.~Rudolph
\nopagebreak
\begin{center}
\parbox{15.5cm}{\sl\samepage
Institut f\"ur Experimentalphysik, Universit\"at Innsbruck, A-6020
Innsbruck, Austria$^{18}$}
\end{center}\end{sloppypar}
\vspace{2mm}
\begin{sloppypar}
\noindent
E.~Bouhova-Thacker,
C.K.~Bowdery,
D.P.~Clarke,
G.~Ellis,
A.J.~Finch,
F.~Foster,
G.~Hughes,
R.W.L.~Jones,
M.R.~Pearson,
N.A.~Robertson,
T.~Sloan,
M.~Smizanska
\nopagebreak
\begin{center}
\parbox{15.5cm}{\sl\samepage
Department of Physics, University of Lancaster, Lancaster LA1 4YB,
United Kingdom$^{10}$}
\end{center}\end{sloppypar}
\vspace{2mm}
\begin{sloppypar}
\noindent
O.~van~der~Aa,
C.~Delaere,$^{28}$
G.Leibenguth,$^{31}$
V.~Lemaitre$^{29}$
\nopagebreak
\begin{center}
\parbox{15.5cm}{\sl\samepage
Institut de Physique Nucl\'eaire, D\'epartement de Physique, Universit\'e Catholique de Louvain, 1348 Louvain-la-Neuve, Belgium}
\end{center}\end{sloppypar}
\vspace{2mm}
\begin{sloppypar}
\noindent
U.~Blumenschein,
F.~H\"olldorfer,
K.~Jakobs,
F.~Kayser,
A.-S.~M\"uller,
B.~Renk,
H.-G.~Sander,
S.~Schmeling,
H.~Wachsmuth,
C.~Zeitnitz,
T.~Ziegler
\nopagebreak
\begin{center}
\parbox{15.5cm}{\sl\samepage
Institut f\"ur Physik, Universit\"at Mainz, D-55099 Mainz, Germany$^{16}$}
\end{center}\end{sloppypar}
\vspace{2mm}
\begin{sloppypar}
\noindent
A.~Bonissent,
P.~Coyle,
C.~Curtil,
A.~Ealet,
D.~Fouchez,
P.~Payre,
A.~Tilquin
\nopagebreak
\begin{center}
\parbox{15.5cm}{\sl\samepage
Centre de Physique des Particules de Marseille, Univ M\'editerran\'ee,
IN$^{2}$P$^{3}$-CNRS, F-13288 Marseille, France}
\end{center}\end{sloppypar}
\vspace{2mm}
\begin{sloppypar}
\noindent
F.~Ragusa
\nopagebreak
\begin{center}
\parbox{15.5cm}{\sl\samepage
Dipartimento di Fisica, Universit\`a di Milano e INFN Sezione di
Milano, I-20133 Milano, Italy.}
\end{center}\end{sloppypar}
\vspace{2mm}
\begin{sloppypar}
\noindent
A.~David,
H.~Dietl,$^{32}$
G.~Ganis,$^{27}$
K.~H\"uttmann,
G.~L\"utjens,
W.~M\"anner$^{32}$,
\mbox{H.-G.~Moser},
R.~Settles,
M.~Villegas,
G.~Wolf
\nopagebreak
\begin{center}
\parbox{15.5cm}{\sl\samepage
Max-Planck-Institut f\"ur Physik, Werner-Heisenberg-Institut,
D-80805 M\"unchen, Germany\footnotemark[16]}
\end{center}\end{sloppypar}
\vspace{2mm}
\begin{sloppypar}
\noindent
J.~Boucrot,
O.~Callot,
M.~Davier,
L.~Duflot,
\mbox{J.-F.~Grivaz},
Ph.~Heusse,
A.~Jacholkowska,$^{6}$
L.~Serin,
\mbox{J.-J.~Veillet}
\nopagebreak
\begin{center}
\parbox{15.5cm}{\sl\samepage
Laboratoire de l'Acc\'el\'erateur Lin\'eaire, Universit\'e de Paris-Sud,
IN$^{2}$P$^{3}$-CNRS, F-91898 Orsay Cedex, France}
\end{center}\end{sloppypar}
\vspace{2mm}
\begin{sloppypar}
\noindent
P.~Azzurri,
G.~Bagliesi,
T.~Boccali,
L.~Fo\`a,
A.~Giammanco,
A.~Giassi,
F.~Ligabue,
A.~Messineo,
F.~Palla,
G.~Sanguinetti,
A.~Sciab\`a,
G.~Sguazzoni,
P.~Spagnolo,
R.~Tenchini,
A.~Venturi,
P.G.~Verdini
\samepage
\begin{center}
\parbox{15.5cm}{\sl\samepage
Dipartimento di Fisica dell'Universit\`a, INFN Sezione di Pisa,
e Scuola Normale Superiore, I-56010 Pisa, Italy}
\end{center}\end{sloppypar}
\vspace{2mm}
\begin{sloppypar}
\noindent
O.~Awunor,
G.A.~Blair,
G.~Cowan,
A.~Garcia-Bellido,
M.G.~Green,
T.~Medcalf,$^{25}$
A.~Misiejuk,
J.A.~Strong,$^{25}$
P.~Teixeira-Dias
\nopagebreak
\begin{center}
\parbox{15.5cm}{\sl\samepage
Department of Physics, Royal Holloway \& Bedford New College,
University of London, Egham, Surrey TW20 OEX, United Kingdom$^{10}$}
\end{center}\end{sloppypar}
\vspace{2mm}
\begin{sloppypar}
\noindent
R.W.~Clifft,
T.R.~Edgecock,
P.R.~Norton,
I.R.~Tomalin,
J.J.~Ward
\nopagebreak
\begin{center}
\parbox{15.5cm}{\sl\samepage
Particle Physics Dept., Rutherford Appleton Laboratory,
Chilton, Didcot, Oxon OX11 OQX, United Kingdom$^{10}$}
\end{center}\end{sloppypar}
\vspace{2mm}
\begin{sloppypar}
\noindent
\mbox{B.~Bloch-Devaux},
D.~Boumediene,
P.~Colas,
B.~Fabbro,
E.~Lan\c{c}on,
\mbox{M.-C.~Lemaire},
E.~Locci,
P.~Perez,
J.~Rander,
B.~Tuchming,
B.~Vallage
\nopagebreak
\begin{center}
\parbox{15.5cm}{\sl\samepage
CEA, DAPNIA/Service de Physique des Particules,
CE-Saclay, F-91191 Gif-sur-Yvette Cedex, France$^{17}$}
\end{center}\end{sloppypar}
\vspace{2mm}
\begin{sloppypar}
\noindent
A.M.~Litke,
G.~Taylor
\nopagebreak
\begin{center}
\parbox{15.5cm}{\sl\samepage
Institute for Particle Physics, University of California at
Santa Cruz, Santa Cruz, CA 95064, USA$^{22}$}
\end{center}\end{sloppypar}
\vspace{2mm}
\begin{sloppypar}
\noindent
C.N.~Booth,
S.~Cartwright,
F.~Combley,$^{25}$
P.N.~Hodgson,
M.~Lehto,
L.F.~Thompson
\nopagebreak
\begin{center}
\parbox{15.5cm}{\sl\samepage
Department of Physics, University of Sheffield, Sheffield S3 7RH,
United Kingdom$^{10}$}
\end{center}\end{sloppypar}
\vspace{2mm}
\begin{sloppypar}
\noindent
A.~B\"ohrer,
S.~Brandt,
C.~Grupen,
J.~Hess,
A.~Ngac,
G.~Prange
\nopagebreak
\begin{center}
\parbox{15.5cm}{\sl\samepage
Fachbereich Physik, Universit\"at Siegen, D-57068 Siegen, Germany$^{16}$}
\end{center}\end{sloppypar}
\vspace{2mm}
\begin{sloppypar}
\noindent
C.~Borean,
G.~Giannini
\nopagebreak
\begin{center}
\parbox{15.5cm}{\sl\samepage
Dipartimento di Fisica, Universit\`a di Trieste e INFN Sezione di Trieste,
I-34127 Trieste, Italy}
\end{center}\end{sloppypar}
\vspace{2mm}
\begin{sloppypar}
\noindent
H.~He,
J.~Putz,
J.~Rothberg
\nopagebreak
\begin{center}
\parbox{15.5cm}{\sl\samepage
Experimental Elementary Particle Physics, University of Washington, Seattle,
WA 98195 U.S.A.}
\end{center}\end{sloppypar}
\vspace{2mm}
\begin{sloppypar}
\noindent
S.R.~Armstrong,
K.~Berkelman,
K.~Cranmer,
D.P.S.~Ferguson,
Y.~Gao,$^{13}$
S.~Gonz\'{a}lez,
O.J.~Hayes,
H.~Hu,
S.~Jin,
J.~Kile,
P.A.~McNamara III,
J.~Nielsen,
Y.B.~Pan,
\mbox{J.H.~von~Wimmersperg-Toeller},
W.~Wiedenmann,
J.~Wu,
Sau~Lan~Wu,
X.~Wu,
G.~Zobernig
\nopagebreak
\begin{center}
\parbox{15.5cm}{\sl\samepage
Department of Physics, University of Wisconsin, Madison, WI 53706,
USA$^{11}$}
\end{center}\end{sloppypar}
\vspace{2mm}
\begin{sloppypar}
\noindent
G.~Dissertori
\nopagebreak
\begin{center}
\parbox{15.5cm}{\sl\samepage
Institute for Particle Physics, ETH H\"onggerberg, 8093 Z\"urich,
Switzerland.}
\end{center}\end{sloppypar}
}
\footnotetext[1]{Also at CERN, 1211 Geneva 23, Switzerland.}
\footnotetext[2]{Now at Fermilab, PO Box 500, MS 352, Batavia, IL 60510, USA}
\footnotetext[3]{Also at Dipartimento di Fisica di Catania and INFN Sezione di
 Catania, 95129 Catania, Italy.}
\footnotetext[4]{Now at University of Florida, Department of Physics, Gainesville, Florida 32611-8440, USA}
\footnotetext[5]{Also IFSI sezione di Torino, INAF, Italy.}
\footnotetext[6]{Also at Groupe d'Astroparticules de Montpellier, Universit\'{e} de Montpellier II, 34095, Montpellier, France.}
\footnotetext[7]{Supported by CICYT, Spain.}
\footnotetext[8]{Supported by the National Science Foundation of China.}
\footnotetext[9]{Supported by the Danish Natural Science Research Council.}
\footnotetext[10]{Supported by the UK Particle Physics and Astronomy Research
Council.}
\footnotetext[11]{Supported by the US Department of Energy, grant
DE-FG0295-ER40896.}
\footnotetext[12]{Now at Departement de Physique Corpusculaire, Universit\'e de
Gen\`eve, 1211 Gen\`eve 4, Switzerland.}
\footnotetext[13]{Also at Department of Physics, Tsinghua University, Beijing, The People's Republic of China.}
\footnotetext[14]{Supported by the Leverhulme Trust.}
\footnotetext[15]{Permanent address: Universitat de Barcelona, 08208 Barcelona,
Spain.}
\footnotetext[16]{Supported by Bundesministerium f\"ur Bildung
und Forschung, Germany.}
\footnotetext[17]{Supported by the Direction des Sciences de la
Mati\`ere, C.E.A.}
\footnotetext[18]{Supported by the Austrian Ministry for Science and Transport.}
\footnotetext[19]{Now at SAP AG, 69185 Walldorf, Germany}
\footnotetext[20]{Now at Groupe d' Astroparticules de Montpellier, Universit\'e de Montpellier II, 34095 Montpellier, France.}
\footnotetext[21]{Now at BNP Paribas, 60325 Frankfurt am Mainz, Germany}
\footnotetext[22]{Supported by the US Department of Energy,
grant DE-FG03-92ER40689.}
\footnotetext[23]{Now at Institut Inter-universitaire des hautes Energies (IIHE), CP 230, Universit\'{e} Libre de Bruxelles, 1050 Bruxelles, Belgique}
\footnotetext[24]{Now at Dipartimento di Fisica e Tecnologie Relative, Universit\`a di Palermo, Palermo, Italy.}
\footnotetext[25]{Deceased.}
\footnotetext[26]{Now at SLAC, Stanford, CA 94309, U.S.A}
\footnotetext[27]{Now at CERN, 1211 Geneva 23, Switzerland}
\footnotetext[28]{Research Fellow of the Belgium FNRS}
\footnotetext[29]{Research Associate of the Belgium FNRS}
\footnotetext[30]{Now at Liverpool University, Liverpool L69 7ZE, United Kingdom}
\footnotetext[31]{Supported by the Federal Office for Scientific, Technical and Cultural Affairs through
the Interuniversity Attraction Pole P5/27}
\footnotetext[32]{Now at Henryk Niewodnicznski Institute of Nuclear Physics, Polish Academy of Sciences, Cracow, Poland}
\setlength{\parskip}{\saveparskip}
\setlength{\topsep}{\savetopsep}
\normalsize
\newpage
\pagestyle{plain}
\setcounter{page}{1}

\section {Introduction}
\label{intro}
\begin{figure}[!b]
\begin{tabular}{cc}
\epsfig{file=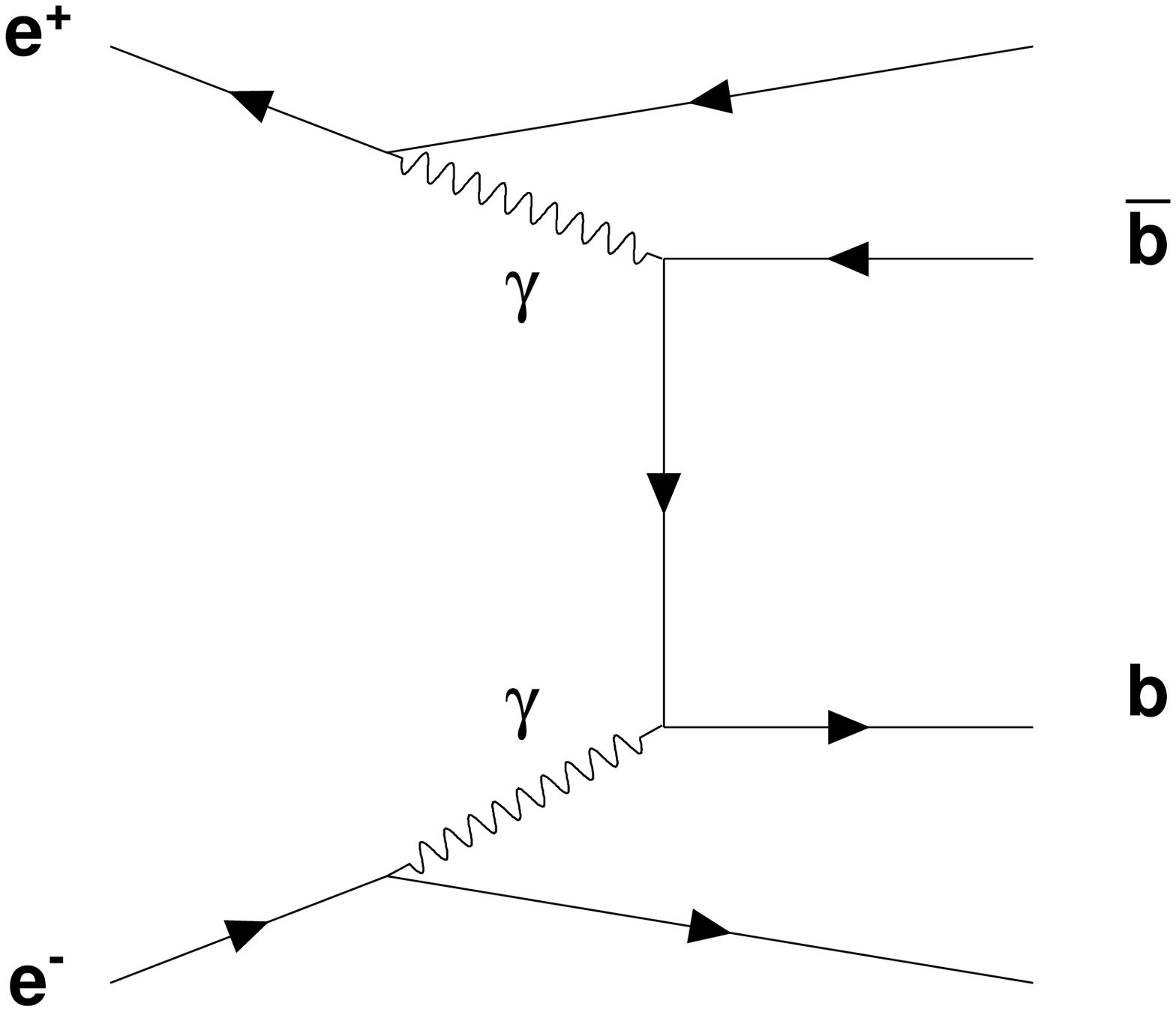,width=0.45\textwidth}
&
\epsfig{file=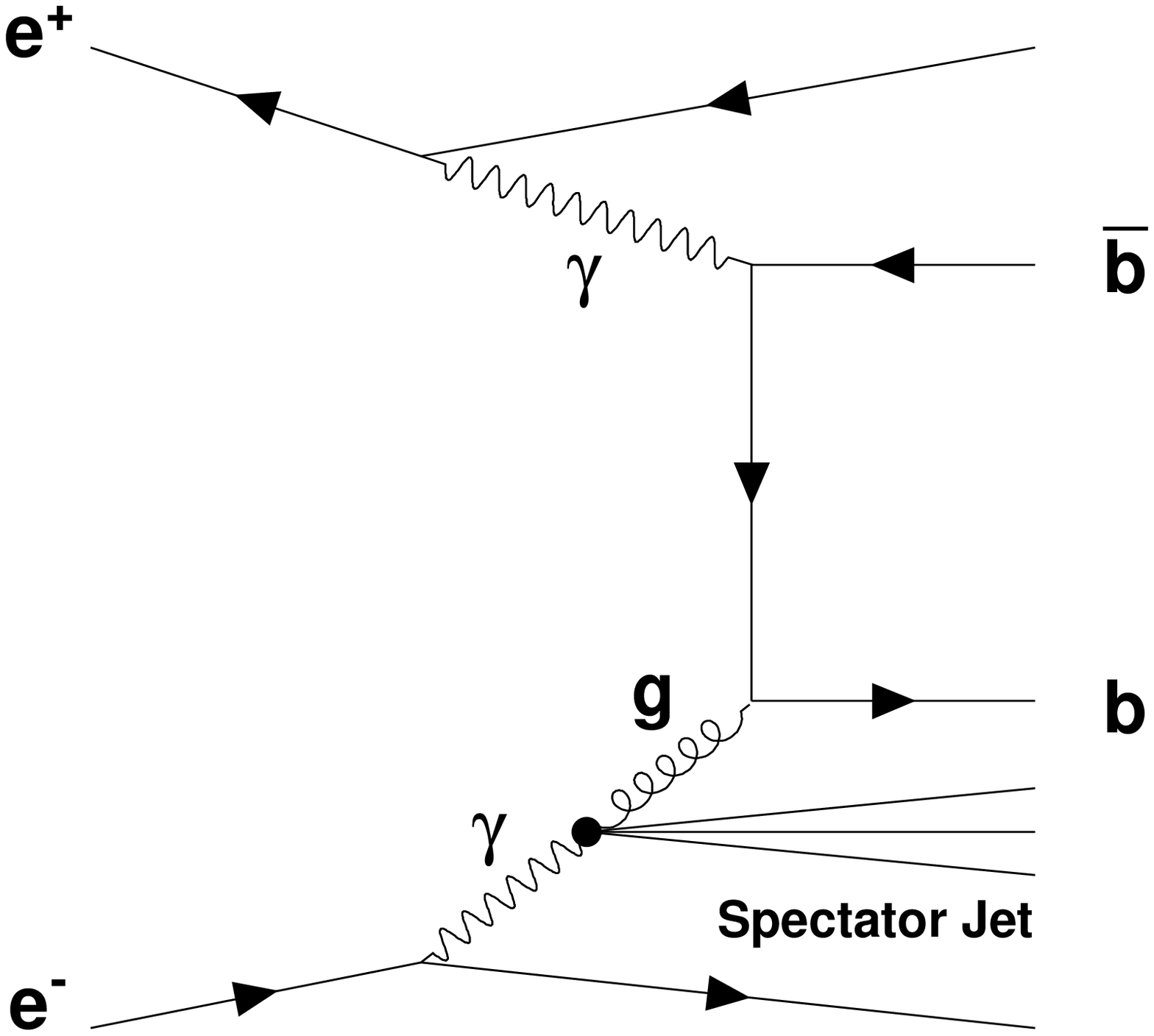,width=0.45\textwidth}
\\
Direct
&
Single Resolved
\\
\end{tabular}
\caption{\label{fig:diagrams}
Diagrams contributing to b-quark production in $\gamma \gamma$
collisions.
}
\end{figure}

The cross section for heavy flavour production in two-photon interactions
is expected to be reliably calculated in perturbative QCD,  particularly in
the case of \beauty-quark production, as the heavy quark
mass introduces a relatively large scale into the process. The cross section  has been
calculated in Next-to-Leading Order (NLO) QCD to be between  2.1 and 4.5 pb~\cite{Drees_et_al}, which is
 two orders of magnitude smaller than that for charm production, which in turn is approximately
6\% of the
total cross section for hadron production. The latter is dominated  by soft
processes involving u,~d and s quarks.
The process of heavy flavour production in two-photon interactions at
LEP energies is dominated by the two classes of diagrams  shown in Fig.~\ref{fig:diagrams}. These are referred to as
the `direct' process in which the photon couples directly to the heavy
quark, and the `single resolved' process in which one photon first
fluctuates into quarks and gluons. This separation is unambiguous up
to
next-to-leading order due to the heavy quark mass~\cite{Frixione}. In the resolved diagram the dominant
process is photon-gluon fusion where a gluon from the resolved photon couples with the
heavy quark. Heavy quark production via  double resolved processes is highly suppressed at LEP
energies~\cite{Drees_et_al}.

The only measurement of \beauty-quark production in two-photon collisions
published to date is by the L3 Collaboration, obtained from a fit to
the transverse momentum of leptons with respect to jets~\cite{L3}: the cross section was measured to be
 about three times the prediction of
NLO QCD. Similar results have been reported at conferences by
OPAL~\cite{OPAL_bb} and DELPHI~\cite{DELPHI_bb}.

This paper presents a measurement of open \beauty-quark production in data
collected  between 1996 and 2000  with an
integrated luminosity of
 $698~\lumi$. During this period the LEP centre of
mass energy ranged from 130 to 209 \GeV, with a mean of 196 GeV. The
result is the first published measurement in which lifetime
information has been used to identify heavy flavour quarks in two-photon physics.
 The paper is organised as follows. Section 2 gives a brief description of the ALEPH
detector, Section 3 presents the event generators used for the simulation of the signal and
backgrounds,  Section 4 describes the jet finding procedure
employed, and Section 5 describes the b tagging procedure.
The initial event selection based on cuts is described in Section 6,
followed in Section 7 by the final  selection which  uses an event weighting procedure.
 In Section 8 the efficiency calculation is described, with the
resulting cross Section given in  Section 9. In
Section 10 the calculation of the  systematic uncertainties is
described, and in Section 11 a number of cross checks are
presented. Finally in Section 12 the final value for the cross section
of open \beauty-quark production is shown.

\section{ALEPH Detector}
\label{sec:Detector}
 The ALEPH detector has been described in detail
elsewhere~\cite{ALEPH}. Critical to this
analysis is the ability to accurately measure charged particles.
 These are detected in a large time projection
chamber (TPC) supplemented by information from the inner tracking
chamber (ITC) which is a cylindrical drift chamber sitting
inside the TPC,
 and a two-layer silicon strip vertex detector
 (VDET) which surrounds the beam pipe close to the interaction point.
The VDET was upgraded in 1996 for the high energy running of LEP.
It consists of 48 modules of double sided silicon strip detectors arranged in two concentric cylinders.
The resolution in $r\phi$ is $10\,\mu{\rm m}$, while that
in $z$ rises from $15\,\mu{\rm m}$ for tracks perpendicular to the
beam direction to 50 $\mu \mathrm{m}$ for tracks at $\cos{\theta}=0.85$~\cite{VDET}.
Charged particle transverse momenta
are measured with a resolution of $\delta p_\mathrm{t} / p_\mathrm{t} = 6 \times
10^{-4}p_\mathrm{t} \oplus 0.005$ ($p_\mathrm{t}$  in $\mathrm{GeV}/c$).

Outside the TPC lies the
electromagnetic calorimeter (ECAL) whose primary purpose is the identification
and measurement of electromagnetic clusters produced by
photons and electrons. It is a lead/proportional-tube sampling
calorimeter segmented in $0.9^{\circ} \times0.9^{\circ} $ projective
towers read out in three sections in depth. It has a total thickness
of 22 radiation lengths and a relative energy resolution of
$0.18/\sqrt{E}\oplus0.009$, ($E$ in GeV) for photons. Outside the ECAL,
a superconducting solenoidal coil produces a $1.5\,{\rm T}$
axial magnetic field and the iron return yoke for the
magnet is instrumented with 23 layers of streamer tubes to form
the hadron calorimeter (HCAL). The HCAL  has a relative energy resolution for
hadrons of $0.85/\sqrt{E}$ ($E$ in GeV). The outermost detector of
ALEPH is a set of muon chambers which consist of two double-layers of
streamer tubes. Near the beam pipe, $3\,{\rm m}$ from the interaction
point on either side, are two luminosity
calorimeters, the LCAL and SiCAL, which are electromagnetic calorimeters
specifically designed to measure the  luminosity via Bhabha scattering.

The information from the tracking detectors and the calorimeters are
combined in an energy flow algorithm~\cite{ALEPH}. For each event, the
algorithm provides a set of charged and neutral reconstructed particles,
called energy-flow objects.

\section{Monte Carlo Simulation}
\label{sec:MC}

The PYTHIA~\cite{Pythia} Monte Carlo program was used to simulate the
two-photon processes. The production of b and c quarks by the direct
and resolved process was modelled separately using PYTHIA 6.1 with
matrix elements including mass effects. For the resolved process the photon's parton
distribution function was the PYTHIA default (SaS 1D)~\cite{SAS}.

 The charm quark production cross section
was normalised using the average of the  measurements made at LEP2,
$\sigma (\mathrm{e^+e^- \rightarrow e^+ e^- c \bar{c}\,X )=   930 \pm 120\,
{\rm pb}}$ \cite{L3,LEP_charm}.
All remaining  hadron production by two-photon collisions
 was simulated using the standard PYTHIA machinery for
incoming photon beams~\cite{Friberg}.
The result of this paper will be compared to a calculation which is
valid for real photons ($Q_\gamma^2 \approx 0$) so events with $Q_\gamma^2 > 6$ were
treated as a background and will be referred to as \tagged events
for the remainder of this paper.
The background from $\mathrm{e^+e^-\rightarrow q\bar{q}}$ was produced using
the KK Monte Carlo program~\cite{KK}.

\section {Jet Finding}
\label{sec:jets}

The direction of partons in an event was estimated using jets found
with a dedicated jet finder (PTCLUS) that optimises the reconstruction of resolved events.
 The PTCLUS algorithm consists of three steps.
\begin{itemize}
\item
 The most energetic energy flow object is
taken as the first jet initiator. The algorithm then loops through all
the remaining objects in order of decreasing energy. If the angle between an object's momentum
vector \boldmath${\it p}$ and the jet momentum $\pclus$ is less than
\unboldmath$90^\circ\,$\boldmath
%
and  the transverse momentum of the object with respect to
$p+\pclus\,\,$\unboldmath
 is smaller
than 0.5 \GeVc\hspace{0.0ex} then, the object is added to the jet.
Otherwise  the object is used as a new jet initiator.
The procedure is repeated until all objects have been assigned to a jet.

\item
The distance between two jets is defined as $Y = M^2/E^2_{\mathrm{vis}}$
where $M$ is  the invariant mass of the pair of jets, assumed to be
massless, and $E_{\mathrm{vis}}$ is the visible energy.
The pair of jets with the smallest value of $Y$ is merged provided  $Y<0.1$
and they are within \unboldmath $90^\circ$ of each other.

\item
The process of merging jets may result in objects having a larger
transverse momentum with respect to the jet to which they have been
assigned than to another jet.
 If this is the case the object is reassigned to the other jet. A
maximum of five reassignments may occur after each merger.

\end{itemize}
The last two steps are repeated until no pair of jets has $Y<0.1$.


\section{b Tagging}
\label{sec:btagging}
This analysis relies on the ALEPH b tagging software developed to
identify b quarks via their long lifetimes~\cite{b-tagging}. It identifies charged tracks
that appear to originate from a point away from the primary event
vertex, and  along the direction of the reconstructed b quark.
The b tagging algorithm relies on the impact parameter of charged
tracks to indicate the presence of long lived particles. The impact
parameter  is defined as the distance of closest approach in space
between a track and the main vertex in the event. It is signed
positive (negative) if the point of closest approach between the track
and the estimated b hadron flight path is in front of (behind) the
main vertex, along the direction of the b momentum estimated using the jets found by PTCLUS.
 The impact
parameter significance $S$ is defined as the signed impact parameter
divided by its estimated measurement error. A fit to the negative $S$
distribution is used to derive a function which when applied to a
single track can be used to obtain $P_{\mathrm{track}}$, the probability that a
track originated at the main event vertex. Only tracks which are
likely to have $S$ reliably measured are used, in particular they are
required to have at least one associated VDET hit. The primary vertex
in an event is found using a procedure specifically designed for use
in b tagging.
Probabilities from
tracks with positive $S$ are combined to form tagging variables. Three
tagging variables are used in this analysis. These are
$P_{\mathrm{event}}$,$P_{\mathrm{jet1}}$, and $P_{\mathrm{jet2}}$
which are respectively the probabilities that the whole event, the
first jet or the second jet contained no decay products from long lived particles.


\section{ Event Selection }
\label{sec:selection}
 The
preselection stage of the analysis identified events which were predominantly
from low $Q_\gamma^2$ two-photon interactions.
Events were required to have
\begin{itemize}
\item at least 5 charged tracks;
\item invariant mass of all energy flow objects (\Wvis) between 4 and 40 \GeVcsq;
\item total energy in the luminosity calorimeters SiCAL and LCAL less than 30 \GeV;
\item  total transverse momentum of the event, relative to the beam
direction, less than 6 \GeVc;
\item thrust less than 0.97.
\end{itemize}

The PTCLUS algorithm was used to find jets using all energy flow objects
with $|\cos\theta|$ less than 0.94. This cut results in the b quark jets
having similar properties in direct and resolved events. Between 1 and 3 jets were
found and ranked by how close their mass was to the nominal b
quark mass of 5 \GeVcsq, with Jet 1 being the closest, Jet
2 the next closest, etc. After the preselection approximately 80\% of the
Jet 1 sample were within $15^\circ$ of a
parton in the direct  $\mathrm{ \gamma \gamma \rightarrow b\bar{b}}$ Monte Carlo,
 while the corresponding figure for the resolved Monte Carlo was $70\%$.

\begin{figure}[!hbt]
\epsfig{file=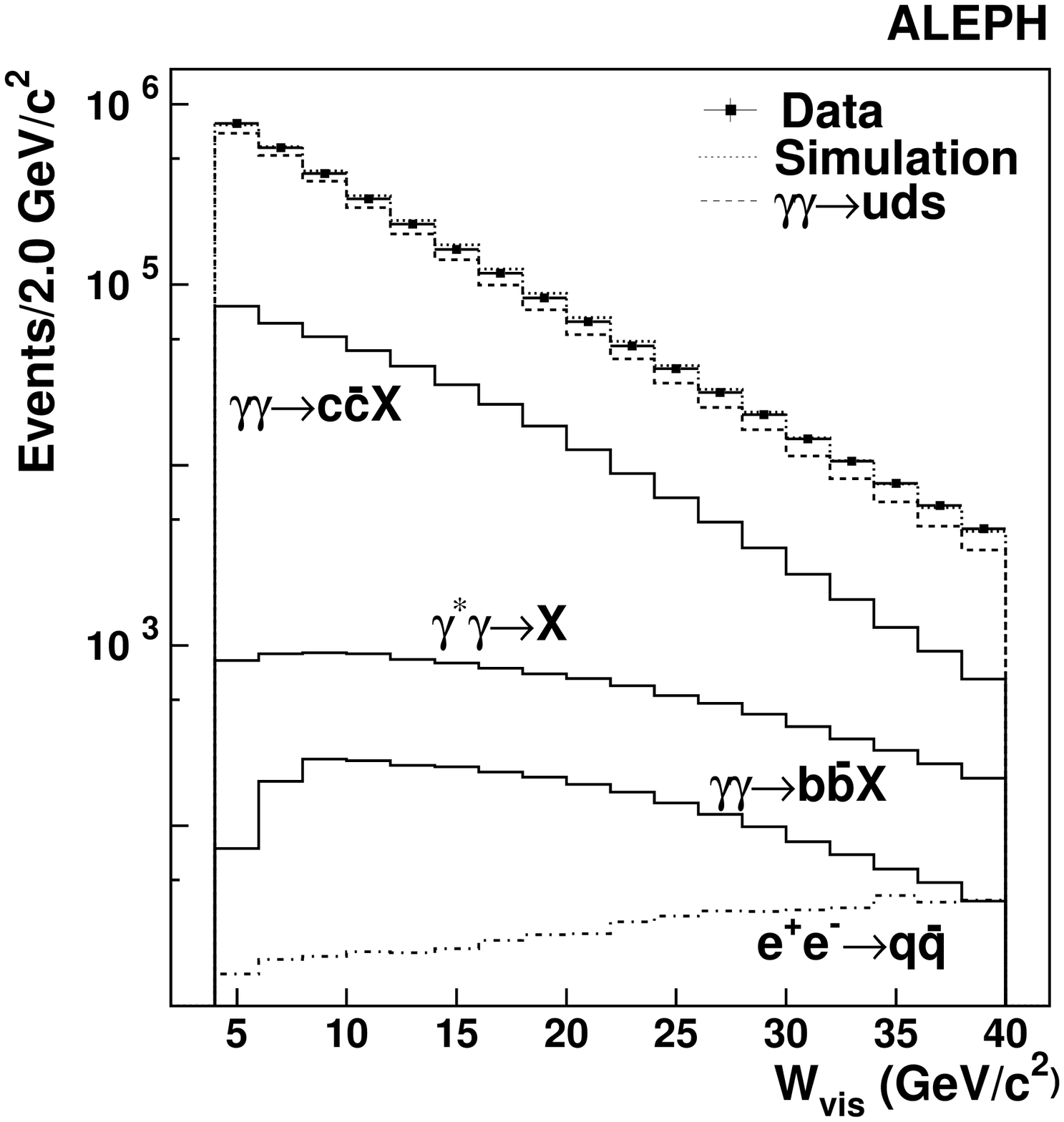,width=0.48\textwidth}
\epsfig{file=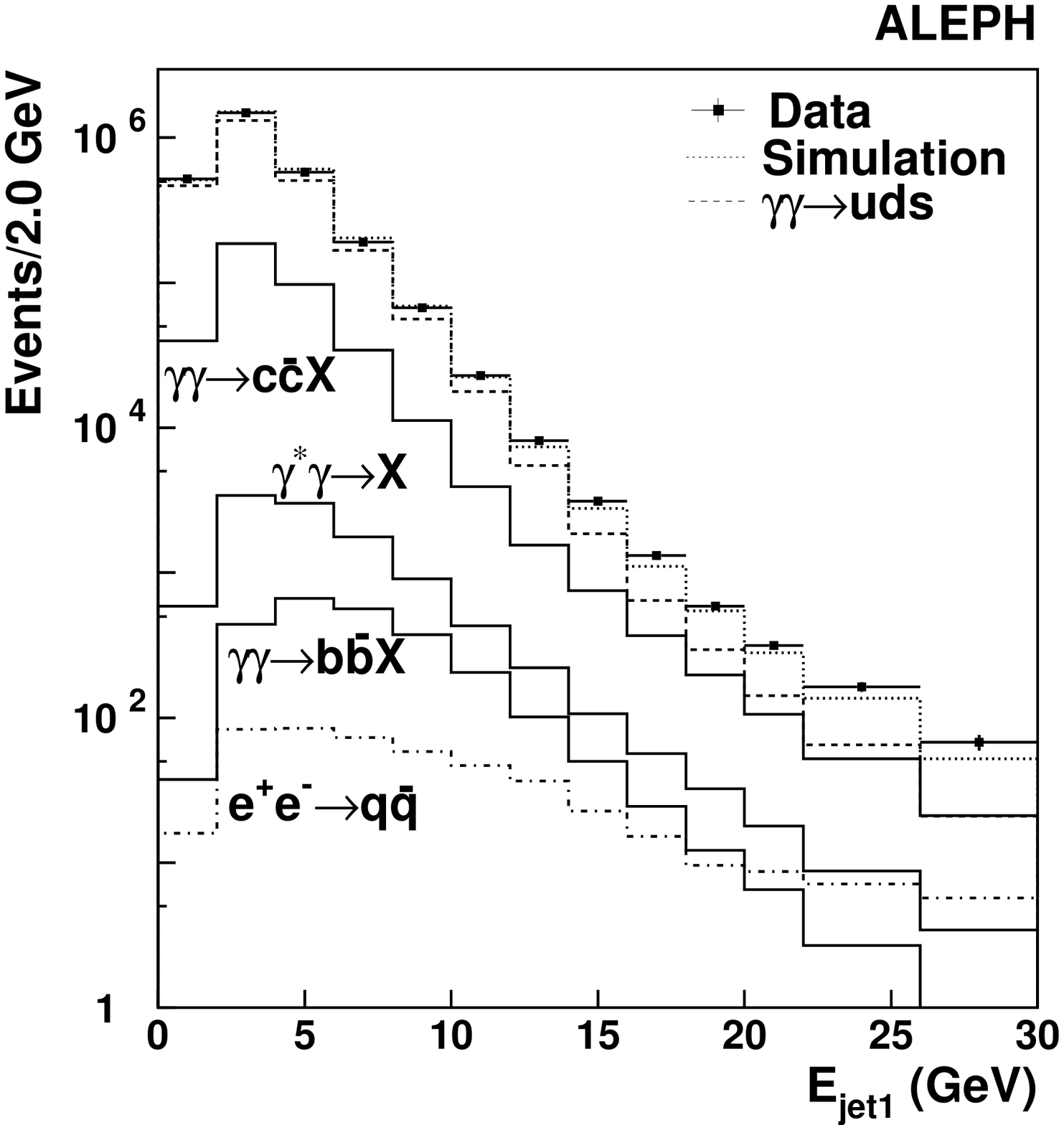,width=0.48\textwidth}
\caption{\label{fig:presel}
Distribution of
\Wvis~ and the energy of Jet 1
in data and simulation after preselection.
}
\end{figure}

 From the distribution of \Wvis, and the energy of Jet 1 shown in Fig.~\ref{fig:presel} it can be
seen that the preselected sample is dominated by events containing light quarks.

 A further selection was applied to enhance the  fraction of events
from the signal process, $\mathrm{ \gamma \gamma \rightarrow b
\bar{b}\,X}$. Events were required to have
\begin{itemize}
\item at least 7 charged tracks;
\item invariant mass of all energy flow objects between 8 and 40 \GeVcsq;
\item      at least two jets;
\item       $P_{\mathrm{event}}<0.05$;
\item    the third  largest impact parameter significance $S$ greater than 0.0;
\item    the fourth largest impact parameter significance $S$ greater than -10.
\end{itemize}

\begin{figure}[bth]
\epsfig{file=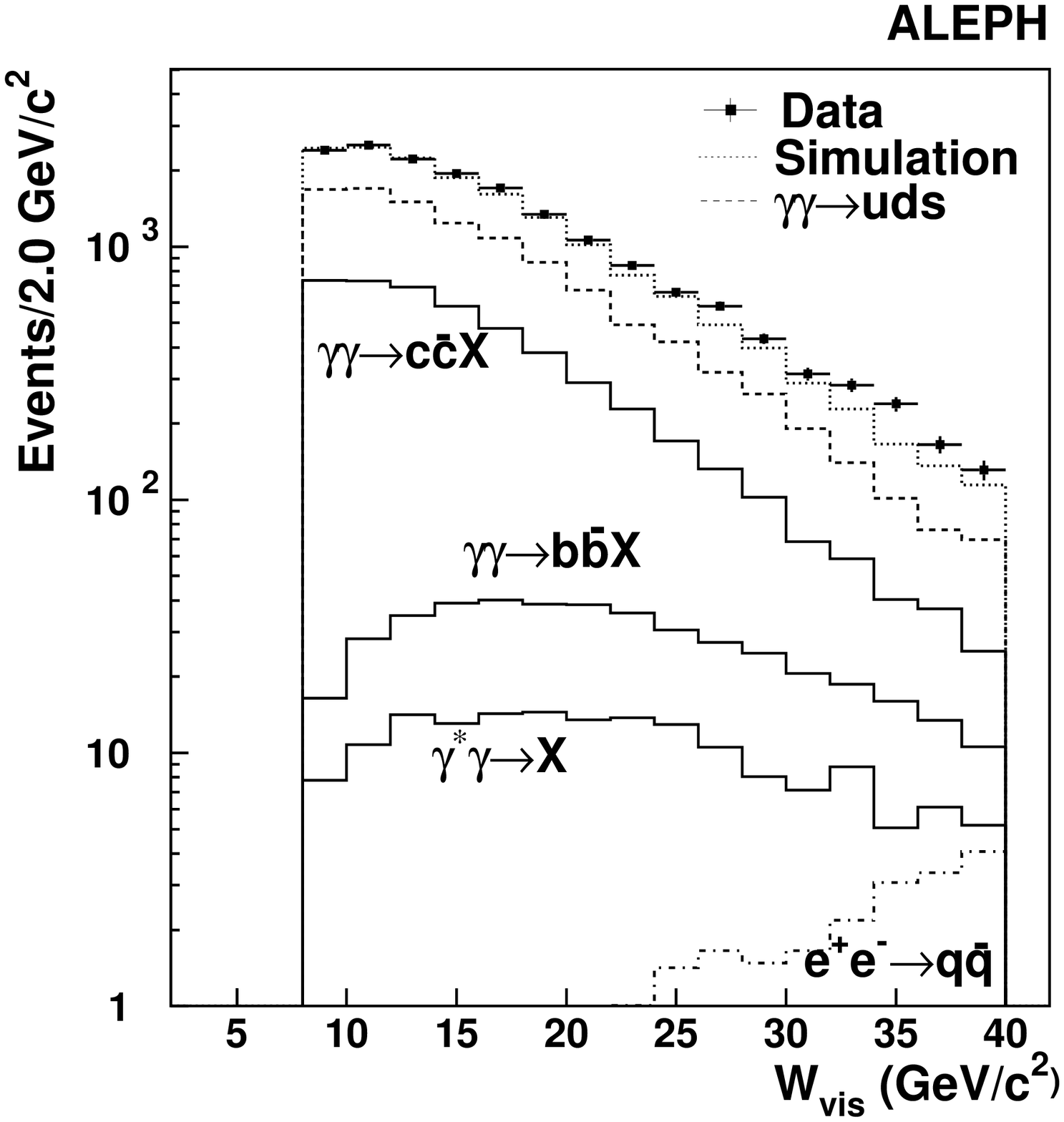,width=0.49\textwidth}
\epsfig{file=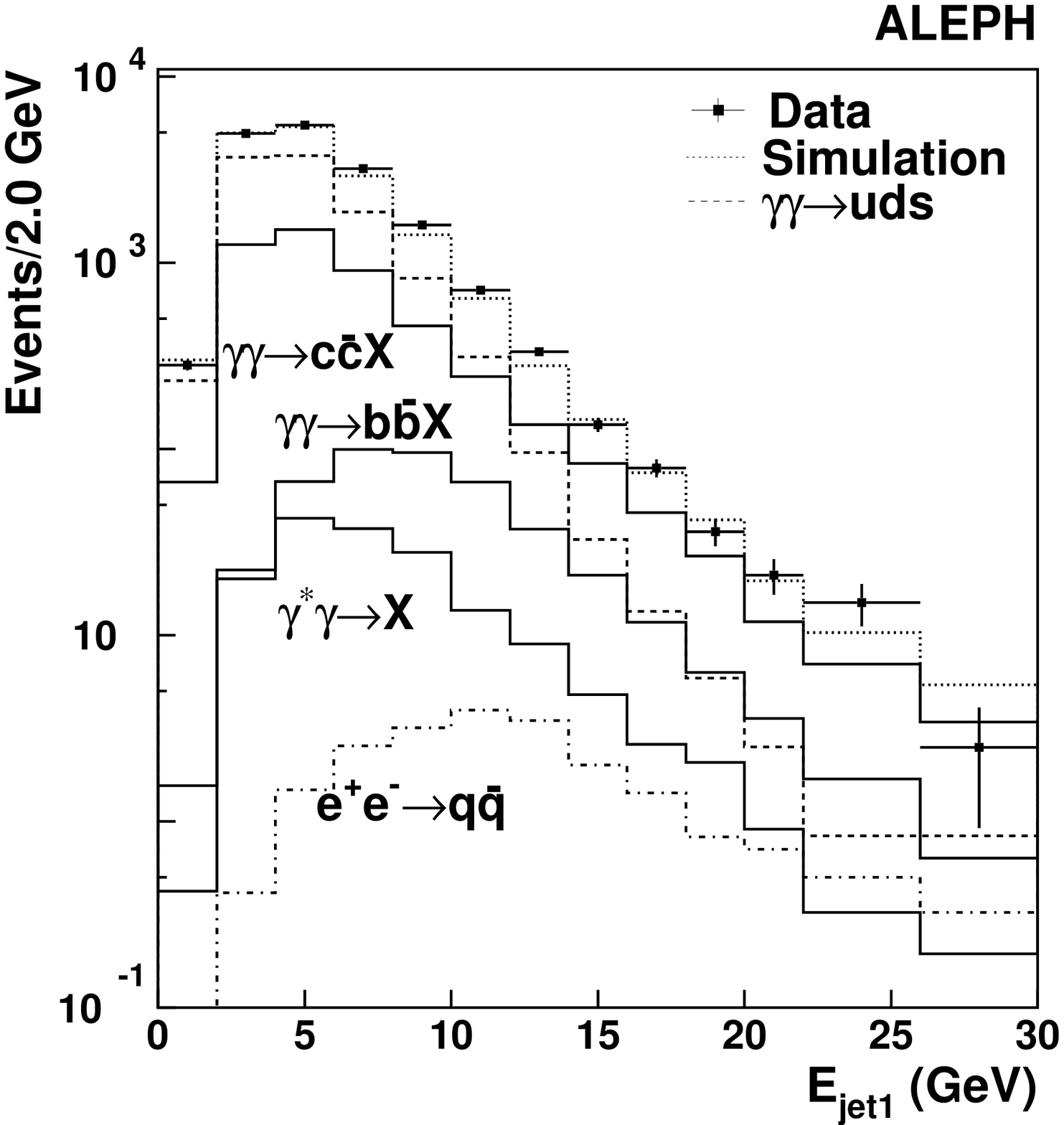,width=0.49\textwidth}
\caption{\label{fig:selection}
Distribution of
\Wvis~ and the energy of Jet 1
in data and simulation after selection.
}
\end{figure}

Figure~\ref{fig:selection} shows the distribution of  \Wvis, and the energy of Jet 1 for
events at this stage of the analysis. Comparison with
Fig.~\ref{fig:presel} shows that while the proportion of the events in
this sample originating from
b quarks has increased compared to the preselection, the
dominant source of events  is still $\mathrm{ \gamma \gamma
\rightarrow uds}$ and $\mathrm{ \gamma \gamma \rightarrow c \bar{c}\,X} $.

\newpage
\section{Iterative Discriminant Analysis}
\label{sec:IDA}

In this analysis the likelihood that an event belongs either to the signal or to the background is
 determined by means of an Iterative Discriminant Analysis (IDA)~\cite{IDA}.
 The details of the method are described in the Appendix.
The method generalises standard linear discriminant  analysis and
proceeds through a series of iterations. At each iteration $i$  events are selected by
applying a cut on the discriminant function for that iteration (\Deye) and
a new discriminant function is then generated for the remaining events.
 The simulated samples described in section~\ref{sec:MC} were used to determine the IDA coefficients.
 A set of 11 variables was chosen as input to the IDA process, these were:
\begin{itemize}
\item  $P_{\mathrm{event}}$,$P_{\mathrm{jet1}}$,$P_{\mathrm{jet2}}$;
\item  mass and transverse momentum of Jet 1;
\item  the five highest track impact parameter significances $S$ seen in the event;
\item  the thrust of the event.
\end{itemize}

After  each IDA iteration the simulations of signal and background were
used to choose whether to perform another iteration, and where to
place the cut on \Deye.
A series of possible values at which to apply a selection  on \Deye~were chosen starting with one that selects 100\% of the
signal and increasing in steps of 1\%  until no signal remained.
At each step the significance of the expected signal above the cut
 was calculated  by dividing it by the predicted error for the
integrated luminosity in the data, including estimated statistical and  systematic
uncertainties. Having determined the value of \Deye~at which the significance was
maximal,  the cut to be applied to the discriminant variable
\Deye~was set at a value $\Delta_\mathnormal{D}$ lower.
 The value of  $\Delta_\mathnormal{D}$ was set to 1.5 for the first iteration,
and halved at each subsequent iteration.
This continued for three iterations
after which there was no further improvement in the predicted
significance.

\begin{figure}[!tbh]
\epsfig{file=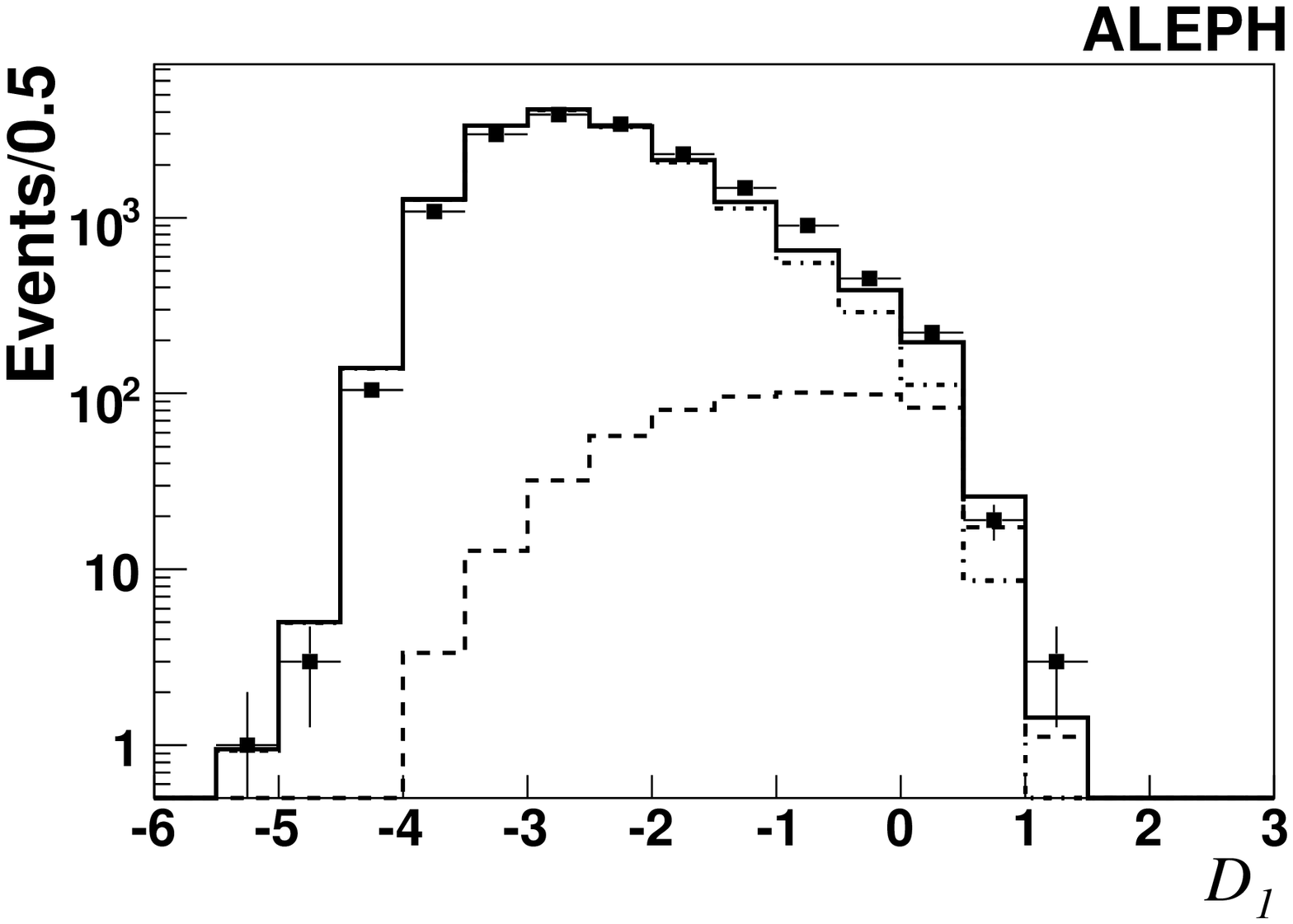,width=0.49\textwidth}
\epsfig{file=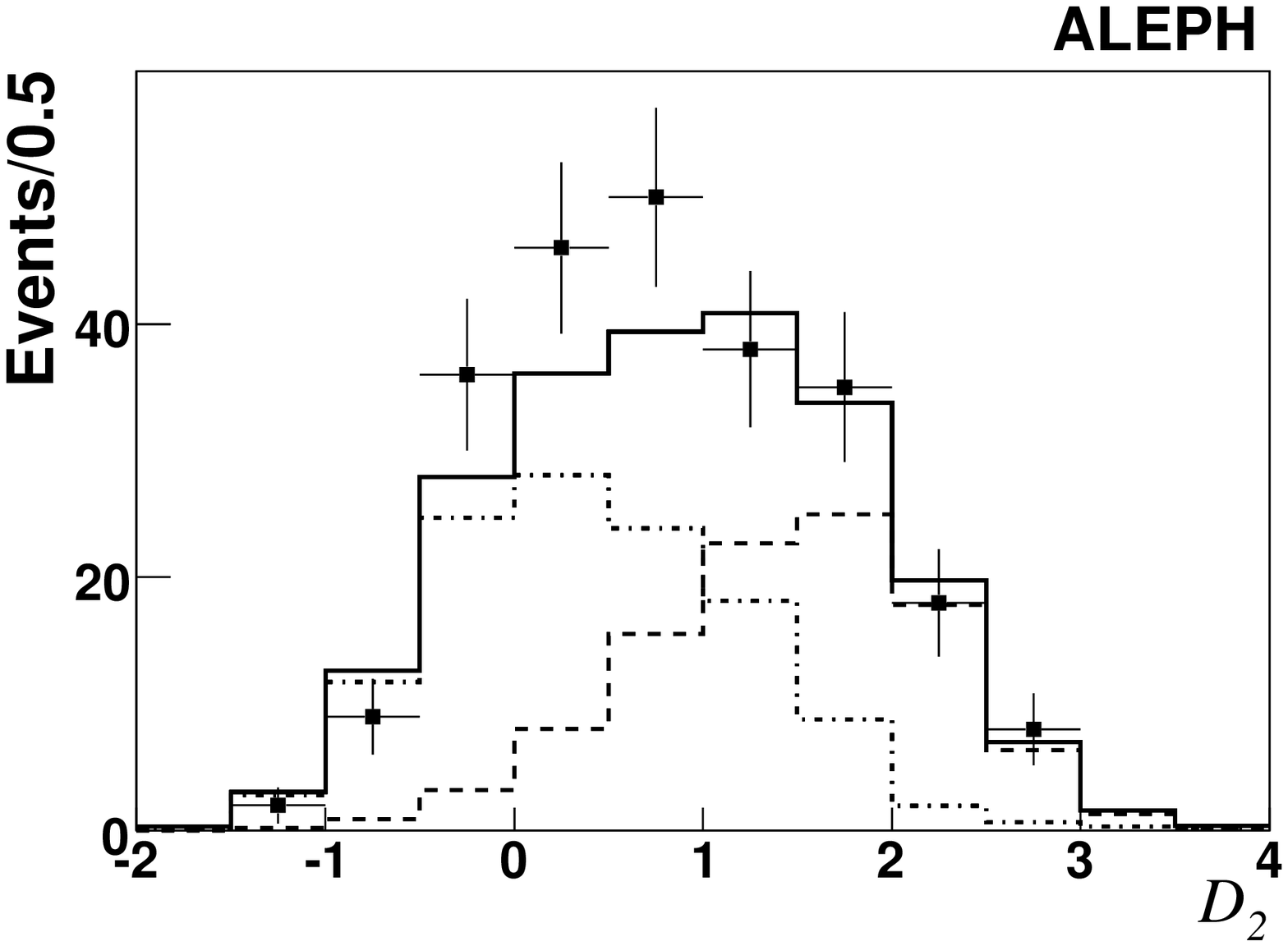,width=0.49\textwidth}
\rule{9em}{0pt}\epsfig{file=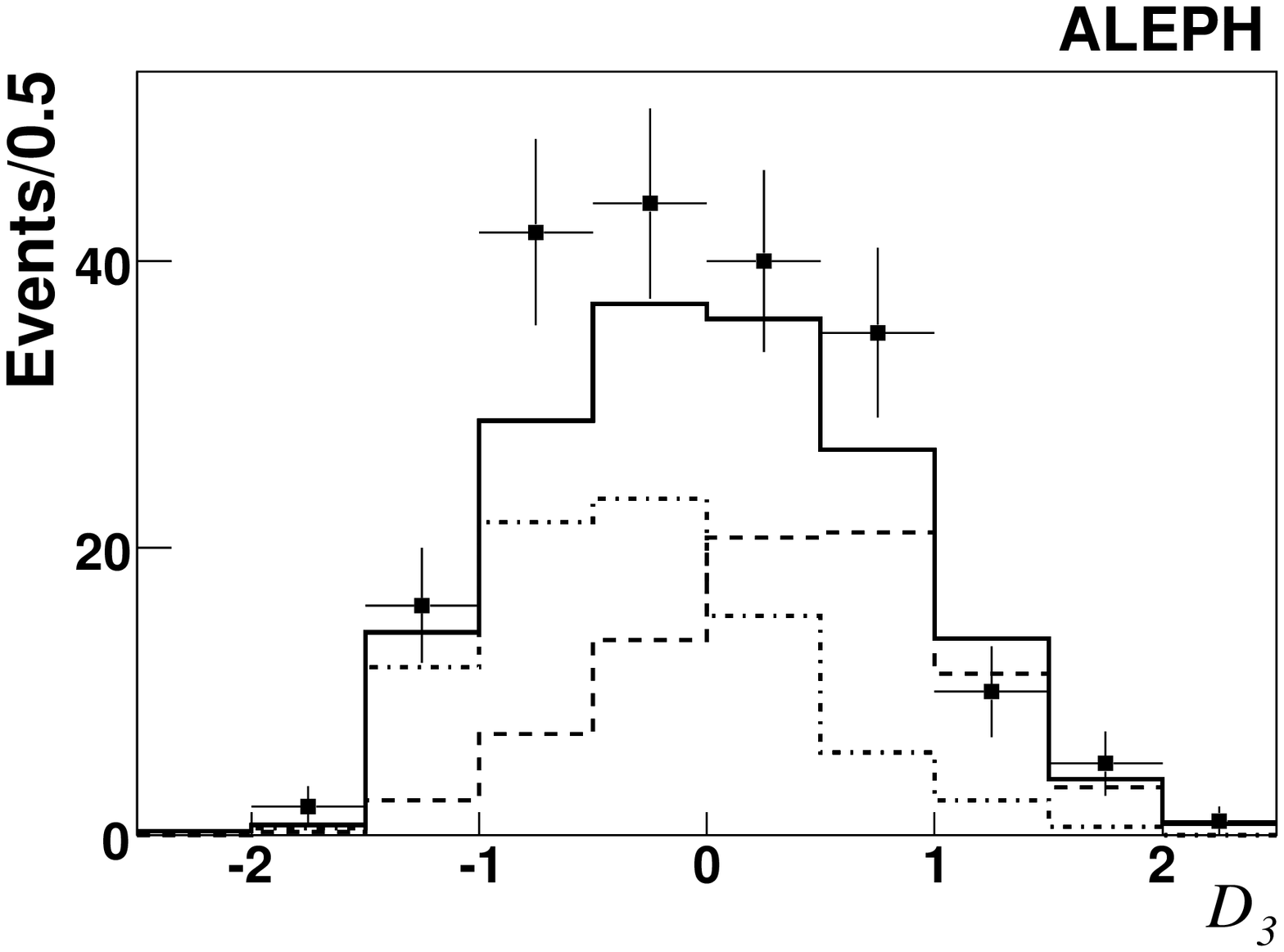,width=0.49\textwidth}
\caption{\label{fig:idaresult}
Distributions of the discriminant variable in data and Monte Carlo samples
after each iteration of the
IDA process. The points with error bars are the data, the dashed
histogram is the simulated signal, the dash-dot histogram is the
simulated background, and
the solid histogram is the sum of signal and background simulations. Each distribution has been translated along
the horizontal axis so that  the selection cut is at zero. The signal
simulation has been weighted according to the fit described in Section~\ref{sec:Effic}.
}
\vspace{-0.15cm}
\end{figure}

 The coefficients of the discriminant analysis and  and cut values
  derived from this
procedure were then applied to the data. However in order to perform various
systematic checks which will be described later, it proved necessary to loosen the
cut on  \Dtwo. This had no significant impact on the purity
of the signal obtained.
The final cut on  \Dthree\, was
chosen to maximise the size of the signal
relative to its uncertainty (both statistical and
systematic). Table~\ref{table:purities} shows the fraction of the total
event sample estimated to come from various sources and the number of
events in the data, at different
stages in the analysis.
 The distribution of the discriminant variables \Deye\hspace{0ex} in the
data and simulation is shown in Fig.~\ref{fig:idaresult} for  each iteration of the IDA process.

The final selection yielded
93 events in the data.
 The background was calculated
using separate samples of simulated events from those used to tune the
IDA parameters. It was found to consist of 18.8 events
from $\mathrm{ \gamma \gamma \rightarrow c \bar{c}\,X}$,
3.9 events from
$ \gamma^* \gamma \rightarrow \,X$
  and
1.5 events from
$\mathrm{e^+ e^-}$ annihilation.

\begin{table}
\begin{center}
 \begin{tabular}{c|c|ccccc}& Cross sect- &
  \multicolumn{5}{c}{ Analysis stage}\\
  \cline{3-7} \raisebox{1.5ex}[0pt]{Sample} & ion (pb)
&1
&2
&3
&4
&5
 \\ \hline
$\mathrm{\gamma \gamma \rightarrow uds}$          & 16000&89 &73 &12 & 9 & 0 \\
$\mathrm{ \gamma \gamma \rightarrow c \bar{c}}\,X$&   930&10 &25 &40 &35 &23 \\
$\gamma^* \gamma \rightarrow X $                  &    84& 0 & 1 & 4 & 5 & 5 \\
$\mathrm{e^+e^-\rightarrow q\bar{q}}$             &    83& 0 & 0 & 2 & 2 & 2 \\
$\mathrm{ \gamma \gamma \rightarrow b \bar{b}}\,X$&     4& 0 & 1 &41 &50 &70 \\
 \hline data & - & 2696021 &  16810 &  244 &  197 &  93 \\
 \end{tabular}
\end{center}
\caption{\label{table:purities}
Summary of the analysis.
The first 5 rows show the cross section used for the simulation
and the  fraction (\%) of each simulated subset at progressive stages of the analysis.
The final row shows the number of events remaining in the data at each
stage.
  The numeric column labels denote
the analysis stages, they are
(1) pre selection, (2) selection, (3) IDA iteration 1, (4) IDA
iteration 2, (5) final cut on \Dthree.}
\end{table}

\clearpage
\section{Efficiency Calculation}
\label{sec:Effic}

The efficiency for signal events to
pass the selection procedure was estimated using a separate
sample of simulated signal events to that used to determine the IDA parameters.
The efficiency is different for the direct and resolved
components so in order to calculate the total efficiency, the
relative size of the two components must be determined. This was found
from the data by performing a fit to the \xgmin
distribution in the data after subtracting the background.
The variable \xgmin is defined as the smallest of
$x_{\gamma}^+$ and $x_{\gamma}^-$ where

\begin{equation}
x_\gamma^\pm=\frac{\sum_{i=1,2}(E^i \pm p_z^i ) }{(E^{\mathrm{tot}} \pm p_z^{\mathrm{tot}})}.
\end{equation}
Here $E^i$, $p_z^i$ are the energy and longitudinal momentum of jet
$i$, while $E^{\mathrm{tot}}$ and $p_z^{\mathrm{tot}}$ are the energy and longitudinal
momentum of the whole event. The sum is calculated for the highest and
second highest energy jets in the event. The $x_\gamma^\pm$ variables
are used in two-photon and photoproduction experiments to distinguish
direct and resolved events. They represent the fraction of the
incoming photon's four-momentum that has gone into the hard scattering
process. For perfectly measured events the value of $x_{\gamma}$ is identically 1
 for direct photons, and less than 1 for
resolved photons, as in the latter case some of the photon's four
momentum is taken away by the spectator jet.
In practice direct events are characterised by having
both $x^+$ and $x^-$ larger than 0.75, while single resolved events
tend to have either $x^+$ or $x^-$ less than 0.75, and double resolved events
have both values less than 0.75~\cite{OPAL_Jets}. In this analysis
only direct and single resolved processes need be considered so \xgmin can be used
to separate them experimentally.

\begin{figure}[!tbh]
\begin{center}
\epsfig{file=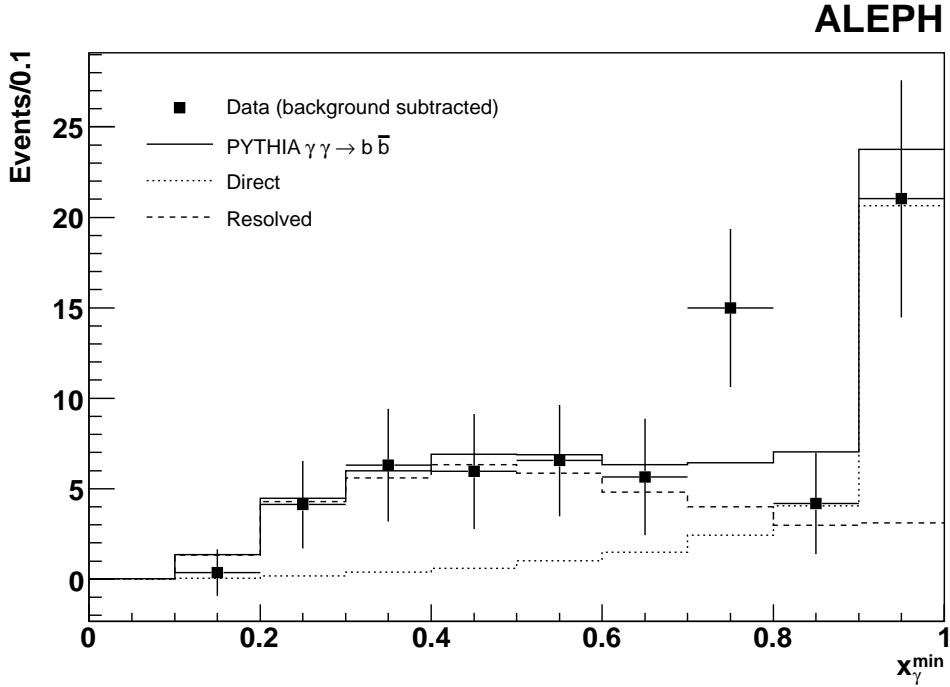,width=0.89\textwidth}
\caption{\label{fig:xgfit}
The \xgmin distribution. The points with error bars are
the data after subtracting the background. The histograms show the
distribution in the simulated direct and resolved signal after
fitting to the data.
}
\end{center}
\end{figure}

The \xgmin distribution is shown in
Fig.~\ref{fig:xgfit} for data after subtracting background and the
simulated direct and resolved components after fitting to the data. The result of the fit is that there are
$30.8
\pm
11.3$ direct
and
 $38.3 \pm 11.9$
resolved events in the data. The efficiencies are
$0.022$ for the  direct term, and
$0.016$ for the resolved term. The mean
efficiency is calculated to be
$0.0184
\pm 0.0009$ where the error comes
from the fit to the fraction of direct and resolved events.
The trigger efficiency for events passing the final cut
 has been measured using independent triggers and found to be negligibly less than 100\%.

\section{Cross Section Calculation}
\label{sec:xsec}

The total cross section is calculated as
\begin{equation}
 \sigma(\mathrm{e^+ e^- \rightarrow e^+ e^- b \bar{b}\, X}) = \frac{N-b}{\epsilon \mathcal{L}}
\end{equation}
where $N$ is the number of events observed, $b$ is the estimated
background, $\epsilon$ is the efficiency and $\mathcal{L}$ is the
luminosity. With $N=
93
$, $b=
24.2$,  $\epsilon=0.0184$ and $\mathcal{L}=698~\mlumi$ the result is $ \sigma(\mathrm{e^+ e^- \rightarrow e^+ e^- b \bar{b}\, X})=
(
5.4
 \pm
0.8
)
\,{\rm pb}$ where the
error is statistical only.

\section{Systematic Uncertainties}
\label{sec:sys}
 \subsection{Background Estimate}
The uncertainty on the background derives from the uncertainty on the
cross section for each component. This is estimated to be 12.5\% for
$\mathrm{ \gamma \gamma \rightarrow c \bar{c}\,X}$~\cite{LEP_charm},
40\% for
  $\gamma^* \gamma \rightarrow X$~\cite{Nissius} and 3\% for
$\mathrm{e^+e^- \rightarrow q \bar{q}}$~\cite{Whalley}. The resulting uncertainty on the
background is
$2.8$ events.

\subsection{Monte Carlo Simulation}

To assess the sensitivity of the efficiency to the
modelling of the physics channels a second sample of signal events was generated using the
HERWIG program~\cite{Herwig} (version 6.201). The difference in
efficiency  obtained using these events was
8.6\%,
and this has been used as a systematic error. The
effect of varying the \beauty-quark fragmentation function in the
simulation was checked and found to be negligible.

\subsection{\boldmath \Wvis~dependence \unboldmath }
Figure~\ref{fig:selection} shows some discrepancy in the \Wvis\hspace{0.1ex}
distribution at the highest values. To check whether this
had any influence on the final result the analysis was repeated with
the maximum accepted \Wvis~set to 30 \GeVcsq. This resulted in the
measured cross section dropping by 0.5 pb. This has been included as a
conservative systematic error.

\section{Cross Checks of the Analysis}
\label{sec:stab}
\subsection{Stability with respect to the \boldmath
\Dthree\hspace{0.0ex} \unboldmath  cut}

\begin{figure}[!bth]
\begin{center}
\epsfig{file=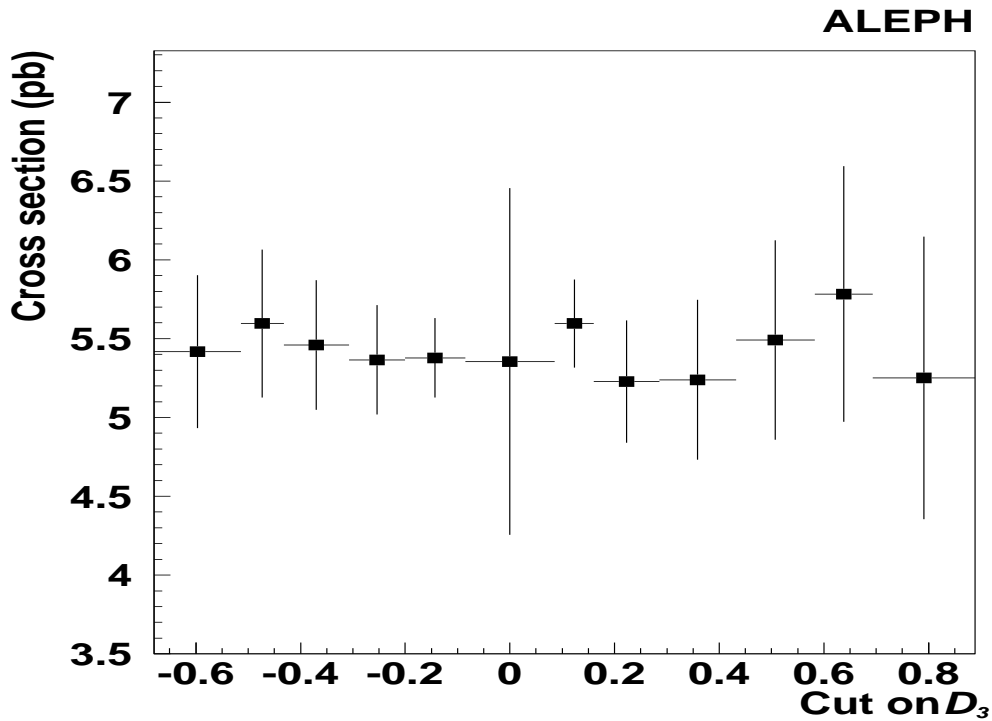,height=10cm,width=0.9\textwidth}
\caption{\label{fig:stability}
Stability of the cross section measurement with respect to changing the
cut on \Dthree. The total error is shown at the chosen cut value ($\mathDthree=0$), while
for the other points the uncertainties relate to the difference of
each point with respect to the chosen cut. The bins are defined such
that each contains 10 more data events than that to its right.}
\end{center}
\end{figure}

An important check of the analysis comes from the dependence
of the result on the \Dthree~cut. In Fig.~\ref{fig:stability} the
cross section measurements obtained when varying the \Dthree~cut either
side of the chosen value are plotted along with the uncorrelated
errors of each point with respect to the point at the chosen cut. No
systematic trend is observed. Similar studies on the \Done~and \Dtwo~cuts
did not reveal significant effects so no additional systematic error
was assigned.

\subsection{\boldmath \Wvis~distribution \unboldmath}
\begin{figure}[tbh]
\begin{center}
\epsfig{file=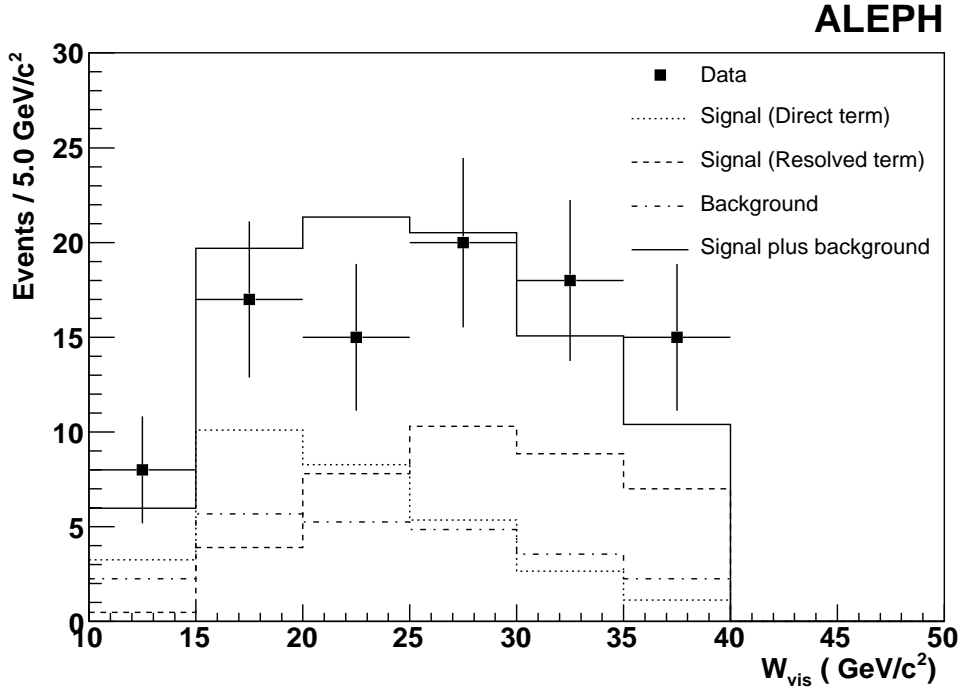,width=0.89\textwidth}

\caption{\label{fig:Wvis}
The distribution of \Wvis~ in selected $
\mathrm{ \gamma \gamma \rightarrow b \bar{b}\,X}$ data. Points with error bars are
the data. The histograms show the
distribution in the background, the direct and resolved signal
and the sum of signal plus background.
}
\end{center}
\end{figure}

An
independent test of the fit to direct and resolved components is given
by the distribution of \Wvis\hspace{0.1ex} which is shown in Fig.~\ref{fig:Wvis}. The direct
 and resolved components also have a significantly different distribution in
 this variable and together they give a good description of the data.

\subsection{Semileptonic decays}

Approximately 20\% of b quarks undergo semileptonic decays, in which an electron or a muon is generated
from the W; therefore about 14 electrons and 14 muons are expected to be produced, on average,
in the observed signal sample of 74 $\mathrm{b \bar{b}}$ events, through direct semileptonic decays.
 Because of the large mass of the b quark, the leptons tend to be at higher transverse momentum relative
to the accompanying jet than those from the decay of the lighter quarks. The production of leptons from
 semileptonic decays of the secondary charm in the b decay chain is also sizeable,
 but the selection efficiency  is considerably smaller because of the softer
momentum and transverse momentum spectra.
All charged tracks with
momentum greater than 2 \GeVc\, were considered as candidate
electrons or muons.

Muons were identified from the pattern of energy deposition left in
the HCAL.
In addition candidate muon tracks were required not be part of a track showing
evidence of a kink in the TPC, to have at least 5 hits in the
ITC, and have a $\mathrm{d}E/\mathrm{d}x$ measurement in the TPC
 consistent with the expectation for a muon.

Electrons were required to have a cluster in the ECAL whose transverse
and longitudinal shape was  consistent with that expected for an
electromagnetic shower, and whose energy was consistent with the
momentum measured in the TPC. In addition they were required to have at least one VDET hit and
at least 3 ITC hits and not be from an identified  converted photon.

Simulation studies show that the majority of
misidentified leptons or leptons not originating from the decay of
b hadrons are found at low transverse momentum relative to the
nearest jet. Requiring the lepton transverse momentum to be greater
than 1 \GeVc\, relative to the nearest jet leaves 0.1\% of misidentified
leptons and 2.5\% from sources other than b hadron decays.

\begin{figure}[tbh]
\begin{center}
\epsfig{file=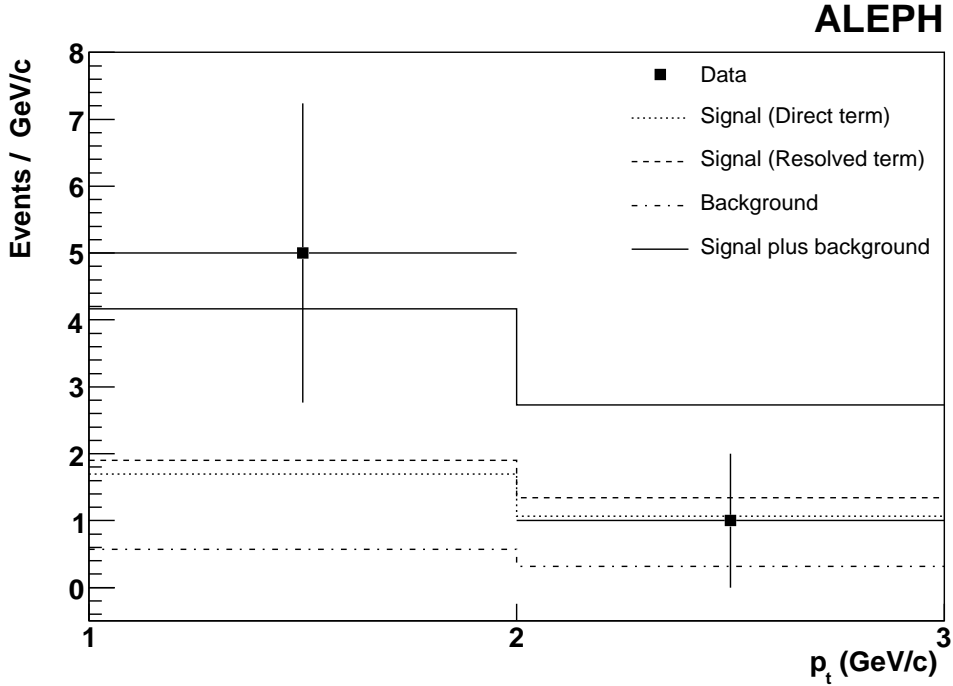,width=0.89\textwidth}

\caption{\label{fig:Leppt}
The transverse momentum of electrons and muons with respect to the
nearest jet in selected $\mathrm{ \gamma \gamma \rightarrow b \bar{b}\,X}$ data.
 Points with error bars are
the data. The histograms show the
distribution in the background, the direct and resolved signal
and the sum of signal plus background.
}
\end{center}
\end{figure}

Figure~\ref{fig:Leppt} shows the distribution of transverse momentum of electrons and muons
with respect to the nearest jet in the
final sample of events. If the lepton is included in the jet then its
momentum has been subtracted from the jet before calculating the
transverse momentum. The signal of 6 leptons is consistent with the
prediction of 6 from  the signal simulation plus
0.9 from the  background.

\section{Conclusions}
\label{sec:conc}
The cross section for the process $\mathrm{e^+e^- \rightarrow e^+e^- b\bar{b}\,X}$
has been measured to be
\[ \mathrm{ \sigma(e^+ e^- \rightarrow e^+ e^- b \bar{b}\, X)  =
(
5.4
\pm
0.8
\,_{stat} \pm
0.8
\,_{syst})\,pb}
\]
which is consistent with the prediction of NLO QCD~\cite{Drees_et_al}
of between 2.1 and 4.5 pb but barely consistent with the result
quoted by the L3 Collaboration~\cite{L3}, $\mathrm{(12.8 \pm 1.7_{stat}      \pm
2.3_{syst}\,)}\, \mathrm{pb}$.

\section*{Acknowledgements}
\label{sec:Acknowledgement}
We would like to thank our colleagues of the accelerator divisions at CERN
 for the outstanding
performance of the LEP machine. Thanks are also due to the many engineers and
technical personnel at CERN and at the home institutes for their contribution
to ALEPH's success. Those of us not from member states wish to thank CERN for
its hospitality.

\newpage
\section*{Appendix A Iterative Discriminant Analysis}
Discriminant analysis is a technique for classifying a set of observations into predefined classes.
 The purpose is to determine the class of an observation based on a set of input variables.
 The model is built based on a set of observations for which the classes are known.
In standard discriminant analysis a set of linear functions of the variables, known as
 discriminant functions, are constructed, such that
$L = \sum_{i=1,n} (b_i x_i) + c$ , where the $b$'s are discriminant coefficients,
 the $x_i$ are the $n$ input variables  and $c$ is a constant. In the
method known as Iterative Discriminant Analysis~\cite{IDA} (IDA) the vector of input
variables $\mathbf x$ is extended to include all their products $x_i x_j $ ( $i \ne
j$). In addition
the process is repeated a number of times with a selection being
applied at each iteration and a new discriminant calculated.
\label{app:ida}
In detail the IDA procedure works as follows:
\begin{itemize}
\item
For each event fill a vector $\mathbf y$ containing the n variables and $(n^2-n)/2$
products of those variables.
\item
Calculate the variance matrix  $\mathbf V = \mathbf V_\mathrm{s}
+\mathbf V_\mathrm{b} $,
where $\mathbf V_\mathrm{s}$ is
the variance matrix of the signal and $\mathbf V_\mathrm{b}$ is the variance matrix of
the background; $\mathbf V_\mathrm{s}$ and $\mathbf V_\mathrm{b}$ are weighted so that they have equal importance.	\unboldmath
\item
Calculate $\mathbf \Delta\mu$, the difference in the means of the signal and background, for each element of $\mathbf y$.
\item
Invert the variance matrix $\mathbf V$ and multiply by $\mathbf
\Delta\mu$, to obtain the vector of coefficients $\mathbf a=\mathbf V^{-1}\mathbf \Delta\mu$.
\item
For each event calculate $D=\mathbf{y^T a y}$.
\end{itemize}
If necessary apply a selection to the events at
some value of $D$ and repeat the procedure as required. The IDA process does not prescribe how such a
 cut should be chosen, or how many iterations should be performed.

\end{document}